\documentclass[prl,twocolumn,showpacs,superscriptaddress,showkeys]{revtex4-1}

\usepackage{amsmath}
\usepackage{amssymb}
\usepackage{chemarr}

\usepackage{hyperref}
\usepackage{braket}

\usepackage{graphicx}

\usepackage{subfigure}
\usepackage{xcolor}

\def\be{\begin{equation}}
\def\ee{\end{equation}}
\def\bea{\begin{eqnarray}}
\def\eea{\end{eqnarray}}

\def\nn{\\ \nonumber}
\renewcommand{\r}[1]{\textcolor{black}{#1}}

\begin{document}


\title{Cell-to-cell information at a feedback-induced bifurcation point}

\author{Amir Erez}
\affiliation{Department of Molecular Biology, Princeton University, Princeton, New Jersey 08544, USA}

\author{Tommy A.\ Byrd}
\affiliation{Department of Physics and Astronomy, Purdue University, West Lafayette, Indiana 47907, USA}

\author{Michael Vennettilli}
\affiliation{Department of Physics and Astronomy, Purdue University, West Lafayette, Indiana 47907, USA}

\author{Andrew Mugler}
\email{amugler@purdue.edu}
\affiliation{Department of Physics and Astronomy, Purdue University, West Lafayette, Indiana 47907, USA}

\begin{abstract}
A ubiquitous way that cells share information is by exchanging molecules. Yet, the fundamental ways that this information exchange is influenced by intracellular dynamics remain unclear. Here we use information theory to investigate a simple model of two interacting cells with internal feedback. We show that cell-to-cell molecule exchange induces a collective two-cell critical point \r{and that the mutual information between the cells peaks at this critical point. Information can remain large far from the critical point on a manifold of cellular states, but scales logarithmically with the correlation time of the system, resulting in an information-correlation time tradeoff. This tradeoff is strictly imposed, suggesting the correlation time as a proxy for the mutual information.}
\end{abstract}

\maketitle

Cells sense and respond to their environment, transforming chemical cues into the modification of signaling molecules, the expression of genes, and the production of proteins. Such signaling networks are often complex, involving, among other features, multiple feedback loops. Yet, the underlying purpose of these networks is to sense and transmit information robustly. For example, in the context of immune response, the complex topology of signaling cascades in T-cells can be such that perturbing a cascade before or after a feedback loop results in dichotomous response \cite{vogel2016dichotomy}. However, coarse-graining the signaling cascade, one can define a basic unimodal-bimodal system, agnostic of biological details, which singles out a particular ``readout'' molecule while integrating out all others. Such coarse graining of the network results in an effective feedback term, which reduces the dynamics to a universal form near a bifurcation point \cite{Munoz2018, Erez2019, Bose2019}. As a result, one can apply critical scaling to these universal dynamics, modified by their non-equilibrium nature \cite{Byrd2019}.

Though powerful, such analysis of intra-cellular dynamics alone treats cells in isolation, equivalent to a very dilute suspension. This ignores the role of cell-to-cell interactions, communicated by means of molecule exchange. Such communication in its simplest form involves only two cells, either similar or different, which produce, degrade, and exchange a molecule. \r{Interaction between two cells is an important biological process, e.g, the immunological synapse \cite{Huppa2003,Daneshpour2019}.} By modeling molecule exchange between two cells, with each cell a generic sense-and-secrete apparatus, one can study the fundamentals of cell-to-cell \r{information}. Investigating the information exchange between two cells in this simple framework is the focus of this work.

\begin{figure}
\begin{center}
\includegraphics[width=1\columnwidth]{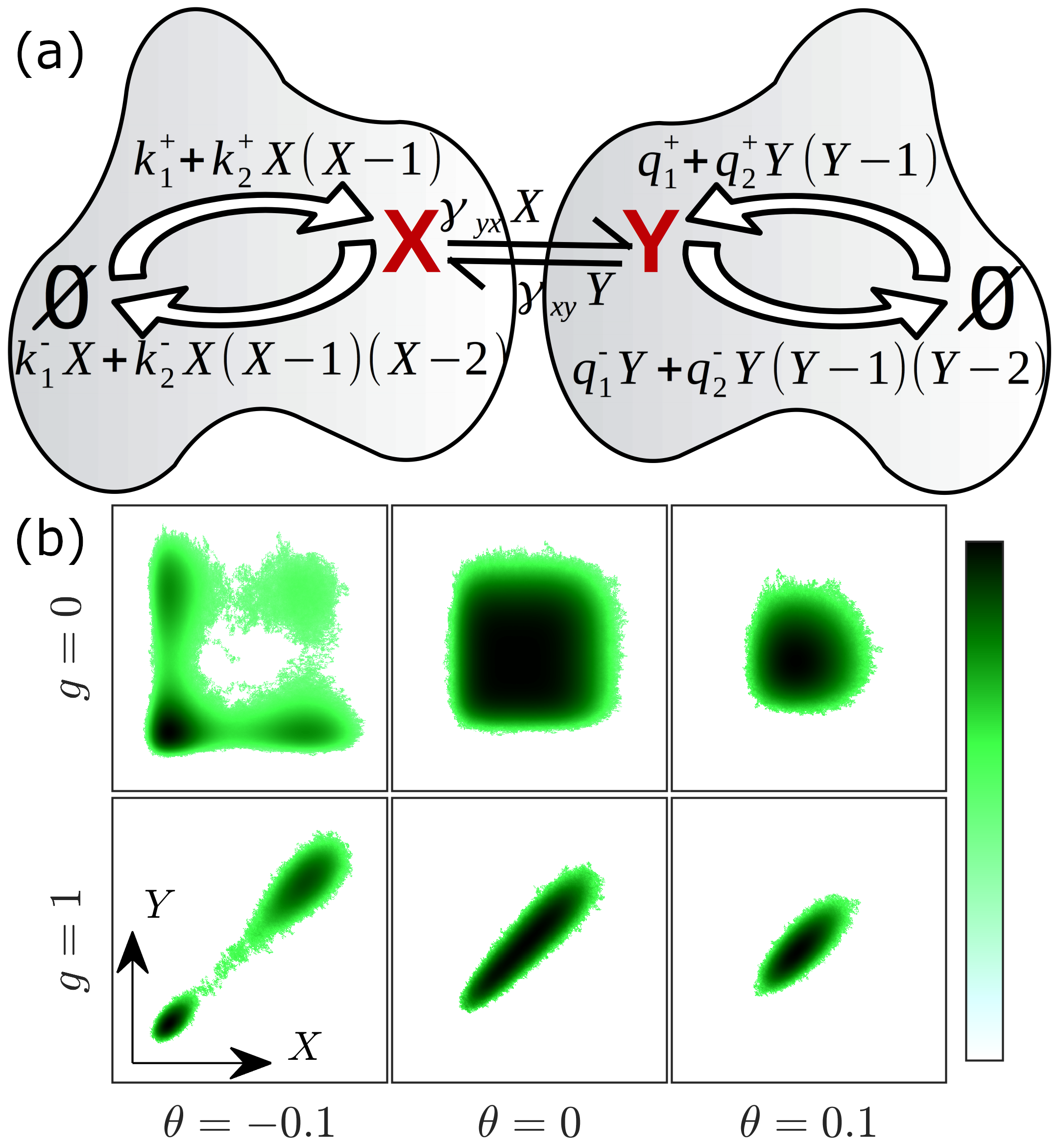}
\end{center}
\caption{\textbf{Model and simulated distributions.} \textbf{(a)} Two-cell Schl\"ogl dynamics with an exchange term $\gamma$. \textbf{(b)} Examples of the joint distribution $P(X,Y)$ from Gillespie simulations with $h=0$ for $g=0$ (top) and $g=1$ (bottom). Color-map corresponds to $\log P$.}
\label{fig1}
\end{figure}

\emph{Model:} Within each cell, biochemical reactions in a complex signaling cascade have the net effect of producing and degrading a molecular species of interest. We specialize to dynamics that can yield either a unimodal or a bimodal molecule number distribution in steady state. Near such a bifurcation point, as was previously shown \cite{Erez2019}, the precise form of the coarse-grained feedback is irrelevant. For convenience we choose to parameterize it using Schl\"ogl's second model \cite{schlogl1972chemical, grassberger1982phase, dewel1977renormalization, nicolis1980systematic, brachet1981critical, prakash1997dynamics, liu2007quadratic, vellela2009stochastic}, a well-studied set of reactions that minimally encodes feedback. Specifically, as illustrated in Fig.~\ref{fig1}(a), in the first (second) cell, species $X$ ($Y$) can be produced spontaneously from bath species at rate $k_1^+$ ($q_1^+$), and can be produced nonlinearly at rate $k_2^+$ ($q_2^+$) via a trimolecular reaction involving two existing $X$ ($Y$) species and a bath species. Species $X$ ($Y$) can be degraded linearly with molecule number at a rate $k_1^-$ ($q_1^-$), or in a reaction involving three existing $X$ ($Y$) molecules at rate $k_2^-$ ($q_2^-$).
In addition to the internal reactions, $X$ ($Y$) can be exchanged from the neighboring cell at rate $\gamma_{xy}$ ($\gamma_{yx}$). Physically, this can be through a gap-junction or through diffusion. Individually, in the absence of exchange, ($\gamma_{xy}=\gamma_{yx}=0$), each of the two constituent cells' molecule number distribution can be either unimodal or bimodal, depending on parameters. If exchange is switched on, ($\gamma_{xy}, \gamma_{yx} > 0$), the system converges to a collective two-cell-state, with the joint distribution unfactorizable \r{in general}, $P(X,Y)\ne P_X(X)P_Y(Y)$.

\emph{Thermodynamic parameters}: Building upon previous work \cite{Erez2019, Byrd2019}, we construct a mapping from Schl\"ogl parameters to Ising-like parameters. Without exchange, the deterministic dynamics corresponding to the reactions in the left cell in Fig.\ \ref{fig1}(a) are $dx/dt = k_1^+ - k_1^-x + k_2^+x^2 - k_2^-x^3$, where we have neglected the small shifts of $-1$ and $-2$ for large $x$. Defining the order parameter $m = (x-n_c)/n_c$, we choose $n_c$ to eliminate the term quadratic in $m$, putting the dynamics in the Landau form \cite{Erez2019}
\begin{equation}
\label{eq:landau}
\frac{dm}{d\tau}=h-\theta m -\frac{m^3}{3},
\end{equation}
where we have defined $n_c = k_2^+/3k_2^-$, $\tau = (k_2^+)^2t/3k_2^-$, $\theta = 3k_1^-k_2^-/(k_2^+)^2 - 1$, and $h = 9k_1^+(k_2^-)^2/(k_2^+)^3 - 3k_1^-k_2^-/(k_2^+)^2 + 2/3$.

\r{The number of molecules in the system is controlled by $n_c$. In fact, $n_c$ controls all scaling properties of the single-cell system, acting as a \emph{finite system size} of the equivalent critical Ising system \cite{Erez2019}. Roughly, in our system, $n_c$ is the value of $X$ or $Y$ at the center of the flat part of the critical distribution ($\theta=0$) in Fig.\ref{fig1}(b).} At \r{small $n_c$, therefore} smaller molecule numbers, small corrections to this mapping can be derived by expanding the known stochastic steady-state distribution around its maximum instead of relying on the deterministic dynamics (\emph{SI Appendix}). We use the corrected mapping in all simulations in this work.

In steady state, $dm/d \tau=0$. We can thus interpret $m$ as an order parameter for the single-cell system, $\theta \equiv (T-T_c)/T_c$ as a reduced ``temperature,'' and $h$ as a dimensionless field. Analogous to the Ising model, when $h=0$ in the single-cell system, $\theta>0$ corresponds to a unimodal steady-state distribution, and $\theta<0$ to a bimodal distribution. Similarly, tuning $h$ biases the distribution to high or low molecule \r{count. The} stochastic steady-state of a single cell at $m,\theta,h=0$ was previously shown to exhibit many properties of equilibrium critical points \cite{Erez2019}. Applying the same mapping to two coupled cells (with $k\to q$ for $Y$) results in the Landau form,
\begin{align}
\frac{dm_X}{d\tau}=h_X-\theta_X m_X - \frac{m_X^3}{3}+g_{XY}m_Y- g_{YX}m_X\,, \\
\frac{dm_Y}{d\tau}=h_Y-\theta_Y m_Y - \frac{m_Y^3}{3}+g_{YX}m_X-g_{XY}m_Y\,,
\label{eq:LandauForm}
\end{align}
where $g_{XY}=3\gamma_{xy}k_2^-/(k_2^+)^2$ and $g_{YX}=3\gamma_{yx}q_2^-/(q_2^+)^2$ parameterize the exchange terms between cells. 

The joint distribution $P(X,Y)$ for \emph{identical} cells is shown in Fig.~\ref{fig1}(b): with identical dimensionless fields ($h_X=h_Y=0$), internal reaction rates ($k_{1,2}^{\pm}=q_{1,2}^{\pm}$), molecular diffusion strengths (\r{$g_{XY}=g_{YX}=g$}), reduced temperatures ($\theta_X=\theta_Y=\theta$) and system size ($n_{c,X}=n_{c,Y}$). In the top row there is no molecular exchange between cells ($g=0$), and each cell is governed solely by its internal dynamics, $P(X,Y)=P_X(X)P_Y(Y)$. Negative $\theta$ yields a \r{\emph{polarized}}, bimodal marginal distribution for each cell, $P_X(X)$ and $P_Y(Y)$. Stochastic fluctuations induce switching between states in each cell individually, resulting in four modes in $P(X,Y)$. When $\theta=0$, each cell sits at its own critical point, resulting in broad and flat marginal distributions, with the joint distribution square-shaped. When $\theta>0$, each cell \r{is \emph{centralized}}, yielding a unimodal marginal distribution \r{about $n_c$}, with the joint distribution also \r{centralized}. In the bottom row of Fig.~\ref{fig1}(b), the effect of \r{molecule exchange} ($g=1$) is evident. When $\theta<0$, each cell is \r{polarized}, and can again access two distinct internal states, but their joint distribution reveals that the cells switch states in concert. When $\theta=0$, each cell can access a broad range of molecule numbers, but \r{exchange} induces the cells to have nearly equal molecule number at all times. This effect is also seen when $\theta>0$, in a smaller, \r{centralized} range of accessible molecule numbers.

\begin{figure}
\includegraphics[width=0.99\columnwidth]{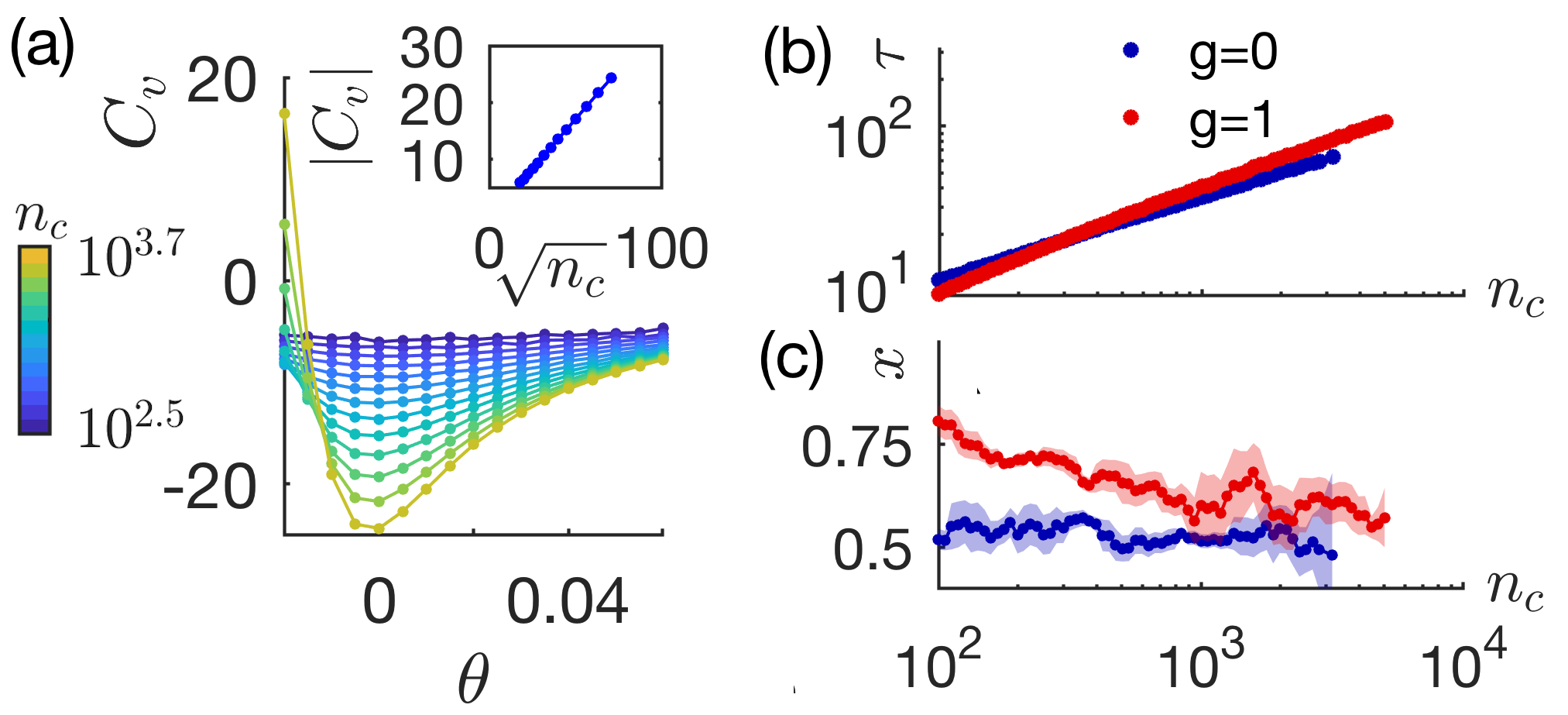}
\caption{\textbf{Scaling of heat capacity \r{and correlation time}}. \textbf{(a)} The heat capacity $C_v$ reaches a minimum at $\theta=0$; Inset - the depth of $C_v(\theta=0)$ scales as $\sqrt{n_c}$. \textbf{(b)} Correlation time $\tau$ at $\theta=0$, dependence on $n_c$, blue: $g=0$ (no \r{exchange}); red: $g=1$. \textbf{(c)} Estimate of the local slope $x = d\ln \tau / d\ln n_c$ from the data in (b), shaded area: \%95 confidence interval.}
\label{fig2}
\end{figure}

Having established that two communicating cells undergo a bifurcation in their collective dynamics at $\theta=0$, we ask:  what are the properties of the two-cell bifurcation point? One can read out the mean-field critical exponents $\beta=1/2$, $\gamma=1$, $\delta=3$ directly from the two-cell Landau form. For the exponent $\alpha$, the single-cell system shows a \emph{minimum} of its heat capacity $C_v$ at $\theta=0$ \cite{Erez2019}, with peak depth depending on the ``system size'' $n_c$. Similarly, for the two-cell system, we calculate $C_v$ directly from the empirical $P(X,Y)$ using $C_v = (1+\theta)\frac{\partial S}{\partial \theta}$ and the Shannon entropy $S = -\sum_{X,Y} P(X,Y) \ln P(X,Y)$. We plot $C_v(\theta,n_c)$ for a range of $n_c$ values in Fig.~\ref{fig2}(a), confirming a minimum of $C_v$ at $\theta=0$, with $C_v(\theta=0) \sim n_c^{1/2}$ (inset). Therefore, at steady state, the two communicating cells near their bifurcation point are in the same \r{static} mean-field universality class as the single-cell system.

\r{When considering the stochastic dynamical system at its bifurcation point in steady state, or its representation as the critical point of two coupled Ising models, an important timescale emerges: the correlation time, $\tau$. The correlation time controls the response of the system to both sudden and gradual changes, a common and important biological scenario, e.g., in the dynamics of response to small-molecule drugs \cite{Byrd2019}.} Fig.~\ref{fig2}(b) shows the dependence of correlation time, $\tau$ on \r{the system size} $n_c$, computed from Gillespie simulations with $\theta=h=0$ using the method of batch means \cite{thompson2010comparison}. The two curves represent a simulation with exchange (red, $g=1$), and without it (blue, $g=0$). To find $x$ in $\tau \sim n_c^{x}$, in Fig.~\ref{fig2}(c) we plot the local slope, $x=d \ln \tau/d \ln n_c$ from Fig.~\ref{fig2}(b). Without exchange, van-Kampen's ``system size'' expansion shows that $x=1/2$ \cite{van1992stochastic,Erez2019}, and this value is confirmed by the blue curve in Fig.~\ref{fig2}(c). With exchange, $x$ is greater than $1/2$ with $x$ tending towards $1/2$ as $n_c$ increases.

\begin{figure}
	\includegraphics[width=1\columnwidth]{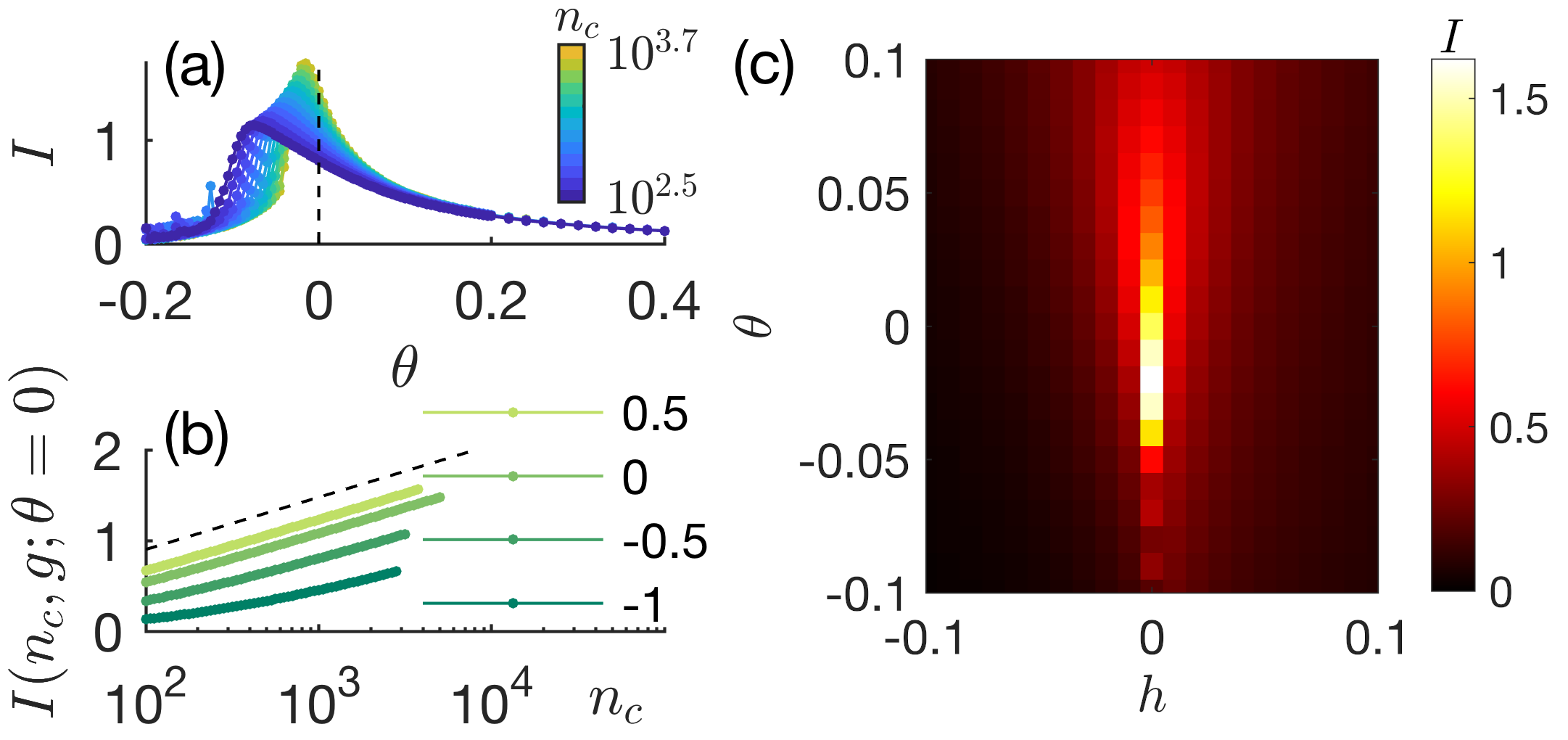}
	\caption{\textbf{Shannon mutual information for identical cells}. \textbf{(a)} Shannon mutual information $I$, for $\theta,h=0$; colors correspond to increasing $n_c$ values from $10^{2.5}$ (blue) to $10^{3.7}$. Dashed horizontal marks $\theta=0$.\textbf{(b)} $I(n_c; \theta=0)$ for different values of $g\in\{10^{-1},10^{-0.5},10^0,10^{0.5}\}$. Dashed black - showing that $I(n_c)= \mbox{const}+ \ln{n_c^{1/4}}$ \textbf{(c)} Heatmap showing $I(\theta,h)$ dependence on both $\theta$ and $h$ with $g=1$ and $n_c=3000$.}
	\label{fig3}
\end{figure}

\begin{figure}[t]
	\includegraphics[width=1\columnwidth]{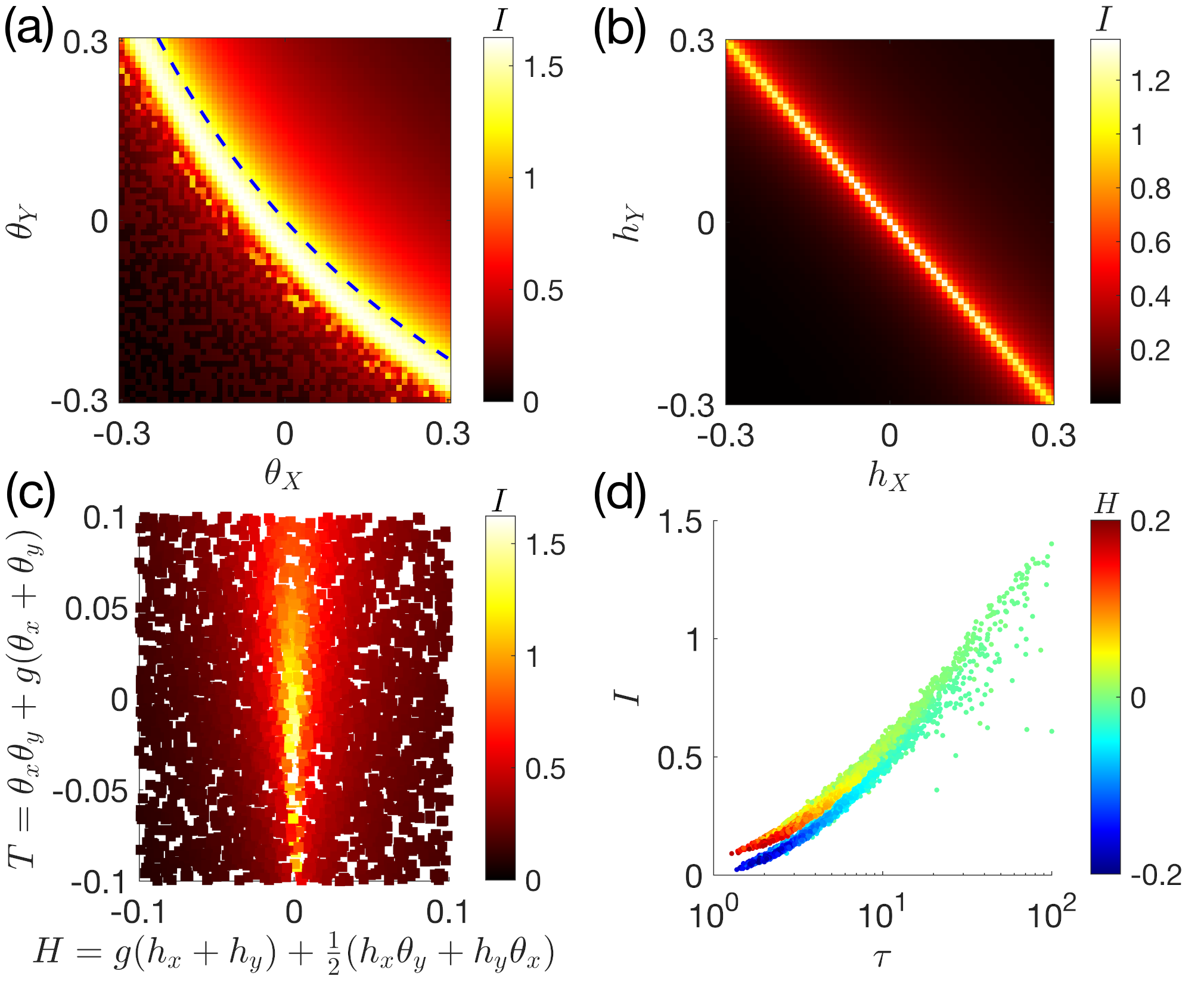}
	\caption{\r{\textbf{Shannon mutual information for dissimilar cells}. \textbf{(a)} Dissimilar $\theta$ values with $h_X=h_Y=0$; blue dashes correspond to $T=0$ from Eq.~\ref{eq:HT}. \textbf{(b)} Dissimilar $h$ values with $\theta_X=\theta_Y=0$. \textbf{(c)} The mutual information $I$ vs.\ $H,T$ (Eq.~\ref{eq:HT}), generated from randomly uniformly drawn $h_x,h_y, \theta_x, \theta_y \in [-0.1, 0.1]$. \textbf{(d)} Mutual information as a function of the correlation time $\tau$ for the data shown in c. In all plots, $g=1$ and $k_1^-=q_1^-=1$ with $n_c=3000$.}}
	\label{fig4}
\end{figure}

In the language of our Ising-like parameters $(h,\theta)$, what values result in the \r{highest} cell-to-cell \r{information}? We quantify \r{information} by means of the Shannon mutual information, $I$, shown in Fig.~\ref{fig3}(a) for identical cells, with $h=0$. Each curve represents a different system size, $n_c$. As $n_c$ increases, $I$ peaks closer to the critical point, $\theta=0$. When $\theta<0$, as shown in Fig.~\ref{fig1}(b), the \r{polarized} bimodal regime inhibits stochastic switching, reducing information exchange. Conversely, when $\theta>0$, noise dominates communcation, suppressing $I$. Moreover, Figure~\ref{fig3}(b) shows that $I(n_c,g, \theta=0) \sim \ln n_c^{1/4}$, to be contrasted with $\tau \sim n_c^{1/2}$, indicating a fundamental trade-off between information and response time in the system: higher precision and faster response-time favor larger and smaller $n_c$, respectively. Fig.~\ref{fig3}(c) shows a heatmap of the mutual information as a function of both $\theta$ and $h$. In addition to again seeing that $I$ is maximal at $\theta = 0$, we also see that moving away from $h=0$ causes $I(\theta,h)$ to sharply decrease. The case $h\ne 0$ biases the baseline production rate, which dampens correlated fluctuations between the two cells and leads to loss of information.

What happens to the information when we relax the requirement for the two cells to be identical? Are there regimes with dissimilar cells that can communicate \r{effectively? Letting $\theta_X\ne \theta_Y$, with $h_X=h_Y=0$, we show in Fig.~\ref{fig4}(a) that $I$ is maximized in a narrow band which crosses $\theta_X=\theta_Y=0$, with $I$ decaying abruptly when $\theta_{X,Y}<0$, but can remain appreciable when $\theta_{X,Y}>0$. This is interesting, because Ising intuition normally proceeds that $T>T_c$ (read $\theta>0$) is disordered and $T<T_c$ is ordered, but here the disordered pair holds higher mutual information farther away from the critical point. The abrupt decay at $\theta<0$ is due to the polarized, bimodal distribution which makes it hard to communicate between the modes. When one cell is polarized ($\theta<0$) and the other is centralized ($\theta>0$), evidently the centralized cell mitigates the polarization when they are correctly matched, resulting in a high information manifold.}

\r{When we let $h_X\ne h_Y$ with $\theta_X=\theta_Y=0$, the mutual information can remain high when $h_X+h_Y\approx 0$, shown in Fig.~\ref{fig4}(b)}, in contrast to the symmetric $h_X=h_Y\ne 0$ case in Fig.~\ref{fig3}(c). The case $h_X+h_Y\approx 0$ models a producer-consumer pair because the field $h$ controls baseline production \cite{Erez2019}. The pair, if rates are matched, \r{shares information effectively.} \r{This is a second special limit of a high information manifold, which we detail in general below. Due to the universal nature of the dynamics near the bifurcation point, both the polarized-centralized pair and the producer-consumer pair show high information in other realizations of the dynamics, such as with Hill-function feedback (\emph{SI} Fig.~\ref{fig:Hill}).} Thus, our simple model captures ubiquitous biological scenarios, showing they support \r{high mutual information}. Analytic results at the Gaussian limit support this observation (\emph{SI Appendix}). 

\r{Having shown that a polarized-centeralized pair and a producer-consumer pair can have high mutual information, two important questions arise: (i) can we explain the high-information manifolds observed in Fig.~\ref{fig4}(a-b) theoretically? (ii) Do these high-information pairings depend on fine-tuning, $h_X=h_Y=0$ for the polarized-centralized pair, and $\theta_X=\theta_Y=0$ for the producer-consumer pair, or are there more realistic high-information pairings that do not depend on setting one pair of cellular parameters to zero? To answer these questions, we first define a set of two-cell collective coordinates.}

\r{Linearizing the deterministic steady state of the Landau form (Eq.~\ref{eq:LandauForm}), we derive the collective coordinates (detailed in the \emph{SI Appendix}),
\bea
	\label{eq:HT}
	T &=& \theta_X\theta_Y+ g(\theta_X+\theta_Y)\,, \nn
	H_X &=& g(h_X+h_Y)+ h_X \theta_Y\,, \nn
	H_Y &=& g(h_X+h_Y)+ h_Y \theta_X\,.
\eea
We further define a symmetric collective field, $H=\frac{1}{2}(H_X + H_Y)$. Note that $H_X=H_Y=0$ is fulfilled when $\theta_X=\theta_Y=0$ and $h_X+h_Y=0$ as in Fig.~\ref{fig4}(b). The case $h_X=h_Y=0$ with $T=0$ is shown in dashed blue in Fig.~\ref{fig4}(a), consistent with the high $I$ contour. The dependence of the mean molecule number as a function of $H,T$ is shown in \emph{SI} Fig.~\ref{fig:MHT}, revealing the characteristic Ising state curves, but here for a two-cell collective state.}

\r{To test that the manifold given by $T=H=0$ in Eq.\ \ref{eq:HT} maintains high information in general, we uniformly draw random configurations of $h_X,h_Y,\theta_X,\theta_Y\in [-0.1, 0.1]$ and plot them on the  $H,T$ axes, colored by the mutual information, shown in Fig.~\ref{fig4}(c). We see a peak at $T=H=0$, confirming our expectation that this manifold implies high mutual information. Thus we extend the simple high mutual information cases, shown in Fig.~\ref{fig4}(a-b) to arbitrary values of $h_X, h_Y, \theta_X,\theta_Y$, ruling out fine-tuning to the critical point as a requirement.} 

\r{By avoiding the critical point, the cells obtain high $I$ near $T=H=0$ but with $\theta_{X,Y},h_{X,Y}\ne 0$. Do they also avoid critical slowing down? We return to the randomly-drawn samples and plot $I$ vs.\ $\tau$ in Fig.~\ref{fig4}(d). Interestingly, we note that all values of $I$ for a set of $\{h_X,h_Y,\theta_X,\theta_Y\}$ collapse on two close branches uniquely determined by the correlation time $\tau$. The branches are distinguished by the sign of $H$, with the lower branch corresponding to $H<0$; this is expected since negative $H$ lowers the mean molecule number, and having fewer molecules to exchange yields less information. The collapse shows that the time/information tradeoff is strictly imposed: there is no ``free lunch'' where the cells can increase $I$ without slowing down. The fact that the mutual information is uniquely defined by $\tau$ is a useful outcome since the correlation time is more readily observed empirically, in contrast to $I$ which requires estimating a joint distribution function.}

\emph{Discussion}: We have shown that coupling two idealized cells, can give rise to a critical system. Extending the Sch\"ogl model, and capitalizing on a mapping between the internal dynamics of each cell and the mean-field Ising model, we cast each constituent of the system in terms of Ising-like quantities. At the collective bifurcation point, $\theta=h=0$, mutual information is maximized, though dynamics are faced with a time/information tradeoff due to critical slowing down. Further, \r{a polarized-centralized pair or} a producer-consumer pair support high mutual information \r{away from the critical point}. \r{We generalize this observation and define a manifold of high mutual information states. However, being away from the critical point does not provide a way to increase information without increasing the correlation time of the system. Rather, the correlation time can serve as a proxy for mutual information in our system.}

The mutual information between two cells, \r{or their correlation time}, can be directly measured from experimental data, such as fluorescence microscopy movies. As such, it is well-suited for high-throughput studies that quantify cellular \r{dynamics} from large-scale biological data-sets. Here, we suggest a minimal model of cell-to-cell communication and with it, a simple theoretical framework. \r{Drawing on the universality of the dynamics near a critical point, the framework does not depend on a particular set of biochemical reactions, though in this manuscript we focused on an extension of Schl\"ogl's second model.} The framework we present could be applied to translate experimental data to thermodynamic and information-theoretic quantities which are informative and interpretable.

\acknowledgments
This work was supported by the Simons Foundation grant 376198 (to A.\ M.). A. E. was supported by the National Science Foundation through the Center for the Physics of Biological Function (PHY-1734030) and by the National Institutes of Health (R01 GM082938).

\bibliographystyle{apsrev}

\newpage
\pagebreak

\appendix

\widetext
\begin{center}
\textbf{\large SI Appendix: Cell-to-cell information at a feedback-induced bifurcation point}
\end{center}
\setcounter{equation}{0}
\setcounter{figure}{0}
\setcounter{table}{0}
\setcounter{page}{1}
\makeatletter
\renewcommand{\theequation}{S\arabic{equation}}
\renewcommand{\thefigure}{S\arabic{figure}}

\section{Date and code availability}
All data and code used for this manuscript are available for download in,

\href{https://github.com/AmirErez/TwoCellInformation/}{https://github.com/AmirErez/TwoCellInformation/}

\section{Ising parameters with stochastic corrections}

The steady state molecule number distribution for the single-cell birth-death process with propensities
\begin{equation}
b_j = k_1^+ + k_2^+j(j-1), \qquad d_j = k_1^-j + k_2^-j(j-1)(j-2)
\end{equation}
as in Fig.\ \ref{fig1}(a) is \cite{Erez2019}
\begin{equation}
p_n = p_0 \prod_{j=0}^{n-1}\frac{b_j}{d_{j+1}},
\end{equation}
with $p_0$ set by normalization. The maximum, or equivalently the maximum of the log
\begin{equation}
\log p_n = \log p_0 + \sum_{j=0}^{n-1} \log b_j - \sum_{j=0}^{n-1} \log d_{j+1},
\end{equation}
occurs when
\begin{equation}
\label{eq:logmax}
0 = \frac{d\log p}{dn} = \log b_{n-1} - \log d_n,
\end{equation}
where we have approximated the sums as integrals. Eq.\ \ref{eq:logmax} implies
\begin{align}
0 &= b_{n-1} - d_n \\
\label{eq:stochdet}
&= k_1^+ + 2k_2^+ - (k_1^- + 3k_2^+ + 2k_2^-)n + (k_2^+ + 3k_2^-)n^2 - k_2^-n^3.
\end{align}
Defining
\begin{align}
\tilde{k}_1^+ &= k_1^+ + 2k_2^+, \\
\tilde{k}_1^- &= k_1^- + 3k_2^+ + 2k_2^-, \\
\tilde{k}_2^+ &= k_2^+ + 3k_2^-, \\
\tilde{k}_2^- &= k_2^-,
\end{align}
we see that Eq.\ \ref{eq:stochdet} is equivalent to the steady state of the deterministic dynamics above Eq.\ \ref{eq:landau} with $k\to\tilde{k}$. Therefore, replacing $k\to\tilde{k}$ in the expressions for $n_c$, $\tau$, $\theta$, $h$, and $g$, and using $m = (n^*-n_c)/n_c$ as the order parameter with $n^*$ the mode(s) of $p_n$, provides a more accurate mapping when molecule numbers are small.

\r{\subsection{Detailed Schl\"ogl to Ising mapping from a deterministic expansion}
\noindent We consider the Schl\"ogl model with birth and death propensities,
\begin{gather}
	b_n = k_1^+ + k_2^+ n(n-1), \qquad d_n = k_1^- n +k_2^- n(n-1)(n-2)\,.
\end{gather}
Coupling two cells, we have
\begin{align}
		\frac{dn_X}{dt} &= k_1^+ + k_2^+(n_1-1)(n_X-2) -k_1^- n_X -k_2^- n_X(n_X-1)(n_X-2) -\gamma_{YX} n_X+\gamma_{XY} n_Y, \\
		\frac{dn_Y}{dt} &= q_1^+ + q_2^+(n_Y-1)(n_Y-2) -q_1^- n_Y -q_2^- n_Y(n_Y-1)(n_Y-2) +\gamma_{YX} n_X-\gamma_{XY} n_Y.
\end{align}
The second derivatives of the two equations vanish at
\begin{equation}
	n_{cX} = 1 + \frac{k_2^+}{3 k_2^-}, 
	\qquad n_{cY} = 1 + \frac{q_2^+}{3 q_2^-}.
\end{equation}
For simplicity, we assume that $n_{cX}=n_{cY}=n_c$, this gives two of the rates in terms of the others and $n_c$
\begin{equation}\label{eq:partialRates}
	k_2^+ = 3k_2^- (n_c-1), \quad q_2^+ =3q_2^- (n_c-1).
\end{equation}
Performing the change of variables $n_X = n_{cX}(m_X+1)$.  Making this substitution gives
\begin{equation}\label{eq:schlm1Unscaled}
	\begin{aligned}
	n_c\frac{dm_X}{dt} &= \left[n_c \left(k_2^- n_c \left(2 n_c-9\right)-\gamma_{YX}+\gamma_{XY}-k_1^- +13 k_2^-\right)+k_1^+-6 k_2^- \right]
	 +\gamma_{XY} n_c m_Y \\
	&+ n_c \left(3 k_2^- \left(n_c-3\right) n_c-\gamma_{YX}-k_1^-+7
	k_2^- \right) m_X - k_2^- n_c^3 m_X^3.
	\end{aligned}
\end{equation}
The analogous equation for $m_Y$ is,
\begin{equation}\label{eq:schlm2Unscaled}
	\begin{aligned}
	n_c\frac{dm_Y}{dt} &= \left[n_c \left(q_2^- n_c \left(2 n_c-9\right)-\gamma_{XY}+\gamma_{YX}-q_1^- +13 q_2^-\right)+q_1^+-6 q_2^- \right]
	+\gamma_{YX} n_c m_X \\
	&+ n_c \left(3 q_2^- \left(n_c-3\right) n_c-\gamma_{XY}-q_1^-+7
	q_2^- \right) m_Y - q_2^- n_c^3 m_Y^3.
	\end{aligned}
\end{equation}
To arrive at the Landau form, we need to rescale time by the factor
\begin{equation}
	f = 3k_2^- n_c^3 \Longrightarrow \tau = \frac{f t}{n_c}.
\end{equation}
Dividing Eq. \ref{eq:schlm1Unscaled} by the factor $f$ gives
\begin{equation}\label{eq:schlm1Scaled}
	\begin{split}
	\frac{dm_X}{d\tau} = \left[ \frac{n_c \left(k_2^- n_c \left(2 n_c-9\right)-\gamma_{YX}+\gamma_{XY}-k_1^-+13 k_2^-\right)+k_1^+-6 k_2^-}{3 k_2^- n_c^3} \right]  \\
	 + \left[ 1-\left(\frac{9 k_2^- n_c+k_1^- -7 k_2^-}{3 k_2^- n_c^2} \right)\right] m_X - \frac{1}{3} m_X^3 \\
	+\frac{1}{3 k_2^- n_c^2} (\gamma_{XY}m_Y-\gamma_{YX}m_X).
	\end{split}
	\end{equation}
	For the $m_Y$ equation, it will be convenient to introduce the ratio of time scales
	\begin{equation}
		\tilde{\rho} = \frac{k_2^-}{q_2^-}.
\end{equation}
The term multiplying $m_Y^3$ in the equation for $dm_Y/d\tau$ is $-f/3\tilde{\rho}$. Multiplying Eq. \ref{eq:schlm2Unscaled} by $\tilde{\rho}/f$ gives
	\begin{equation}\label{eq:schlm2Scaled}
	\begin{split}
	\tilde{\rho}\frac{dm_Y}{d\tau} = \left[ \frac{n_c \left(q_2^- n_c \left(2 n_c-9\right)-\gamma_{XY}+\gamma_{YX}-q_1^-+13 q_2^-\right)+q_1^+-6 q_2^-}{3 q_2^- n_c^3} \right]  \\
	+ \left[ 1-\left(\frac{9 q_2^- n_c+q_1^- -7 q_2^-}{3 q_2^- n_c^2} \right)\right] m_X - \frac{1}{3} m_X^3 \\
	+\frac{1}{3 q_2^- n_c^2} (\gamma_{YX} m_X-\gamma_{XY} m_Y)\,.
	\end{split}
\end{equation}
Anticipating the Landau form, we define the following parameters,
\begin{equation}\label{eq:schlparams}
	\begin{gathered}
	h_X = \left[ \frac{n_c \left(k_2^- n_c \left(2 n_c-9\right)-\gamma_{YX}+\gamma_{XY}-k_1^-+13 k_2^-\right)+k_1^+-6 k_2^-}{3 k_2^- n_c^3} \right], \\
	h_Y = \left[ \frac{n_c \left(q_2^- n_c \left(2 n_c-9\right)-\gamma_{XY}+\gamma_{YX}-q_1^-+13 q_2^-\right)+q_1^+-6 q_2^-}{3 q_2^- n_c^3} \right], \\
	\theta_X = \left(\frac{9 k_2^- n_c+k_1^- -7 k_2^-}{3 k_2^- n_c^2} \right)-1, \\
	\theta_Y = \left(\frac{9 q_2^- n_c+q_1^- -7 q_2^-}{3 q_2^- n_c^2} \right)-1, \\
	 g_{XY} =\frac{\gamma_{XY}}{3 k_2^- n_c^2}, \qquad g_{YX} =\frac{\gamma_{YX}}{3 k_2^- n_c^2}\,.
	\end{gathered}
\end{equation}
These allow us to write the equations as,
\begin{equation}
	\begin{gathered}
	\frac{dm_X}{d\tau} = h_X -\theta_X m_X -\frac{1}{3}m_X^3 -g_{YX}m_X + g_{XY}m_Y, \\
	\tilde{\rho}\frac{dm_Y}{d\tau} = h_Y -\theta_Y m_Y -\frac{1}{3}m_Y^3 +\tilde{\rho}g_{YX}m_X - \tilde{\rho}g_{XY}m_Y.
	\end{gathered}
\end{equation}	
Now we invert the expression in Eq. \ref{eq:schlparams}, making the simplification that $g_{XY}=g_{YX}=g$ to find,
\begin{equation}
	\begin{gathered}
		k_1^+ = \frac{n_c^3 \left(3(h_X+\theta_X)+1\right)-6 n_c+6}{3 n_c \left(\left(\theta_X+1\right) n_c-3\right)+7}\,k_1^-\,, \\
		q_1^+ = \frac{n_c^3 \left(3 (h_Y+\theta_Y)+1\right)-6 n_c+6}{3 n_c \left(\left(\theta_Y+1\right)	n_c-3\right)+7}\,q_1^-\,, \\
		k_2^- = \frac{k_1^-}{3 n_c \left(\left(\theta_X+1\right) n_c-3\right)+7}\,, \\
		q_2^- = \frac{q_1^-}{3 n_c \left(\left(\theta_Y+1\right) n_c-3\right)+7}\,, \\
		k_2^+ = 3k_2^- (n_c-1)\,, \\
		q_2^+ = 3q_2^- (n_c-1)\,, \\
		\gamma_{YX} = \frac{3 n_c^2}{3 n_c \left(\left(\theta_X+1\right) n_c-3\right)+7}\,g k_1^- =\gamma_{XY}\,.
	\end{gathered}
\end{equation}
Importantly, the canonical Landau form (Eq.~\ref{eq:LandauForm}) requires that $\tilde{\rho}=1$. This dictates a relation between the degradation timescales $k_1^-$ and $q_1^-$ such that,
\be
\frac{q_1^-}{k_1^-} = \frac{3 n_c \left(\left(\theta_Y+1\right) n_c-3\right)+7}{3 n_c \left(\left(\theta_X+1\right) n_c-3\right)+7}
\ee
}

\r{\section{Derivation of the collective coordinates $T,H$}
\noindent Before deriving collective coordinates for the two-cell system, let us first consider a single cell. The single-cell Landau dynamics are, 
\be
\frac{dm}{dt} = h - \theta m - m^3/3\,.
\ee
We can discover the bifurcation point by dropping the cubic term and considering steady state. Specifically, we have $m^* = h/\theta$ in steady state. This steady state is consistent with the following requirements: (i) when the numerator $h=0$ there is no bias and so $m^*=0$. (ii) As the system parameters are taken to the bifurcation point, $\theta \rightarrow 0$, the denominator enhances fluctuations. Indeed, a vanishing denominator makes m* infinitely susceptible to changes in the bias $h$. We proceed to similarly derive collective two-cell coordinates.}

\r{At steady state, dropping the cubic term from the two-cell Landau form (Eq.\ref{eq:LandauForm}) gives,
\be
\left(\begin{array}{cc}
\theta_X+g	& -g \\
-g			& \theta_Y+g
\end{array} \right) 
\left(\begin{array}{c} m_X \\
					   m_Y \end{array}\right) = 
\left(\begin{array}{c} h_X \\
					   h_Y \end{array}\right)
\ee
Multiplying on the left by the inverse matrix yields,
\begin{equation*}
	\left(\begin{array}{c} 
		m_X \\
		m_Y 
	\end{array} \right)
	=
	\frac{1}{g(\theta_X+\theta_Y)+\theta_X\theta_Y}
	\left(\begin{array}{c} 
		g(h_X+h_Y)+\theta_Y h_X\\
		g(h_X+h_Y)+\theta_X h_Y
	\end{array} \right)
	=
	\frac{1}{T} \left(\begin{array}{c} 
		H_X\\
		H_Y
	\end{array} \right)
\end{equation*}
The last equality defines the collective coordinates $H_X, H_Y, T$ as used in Eq.~\ref{eq:HT} in the main text.
}

\begin{figure}[!h]
	\includegraphics[width=0.25\columnwidth]{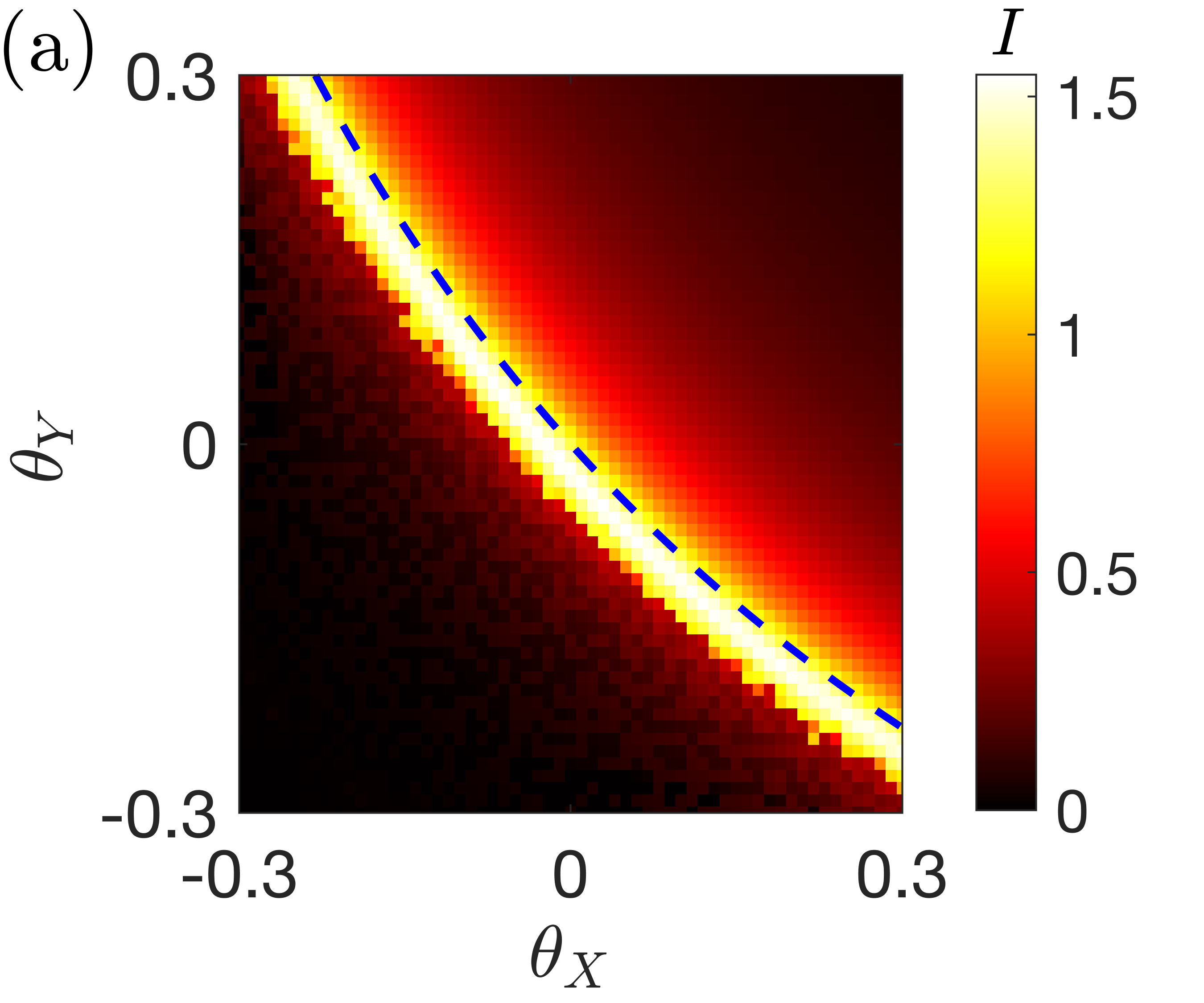}
	\includegraphics[width=0.25\columnwidth]{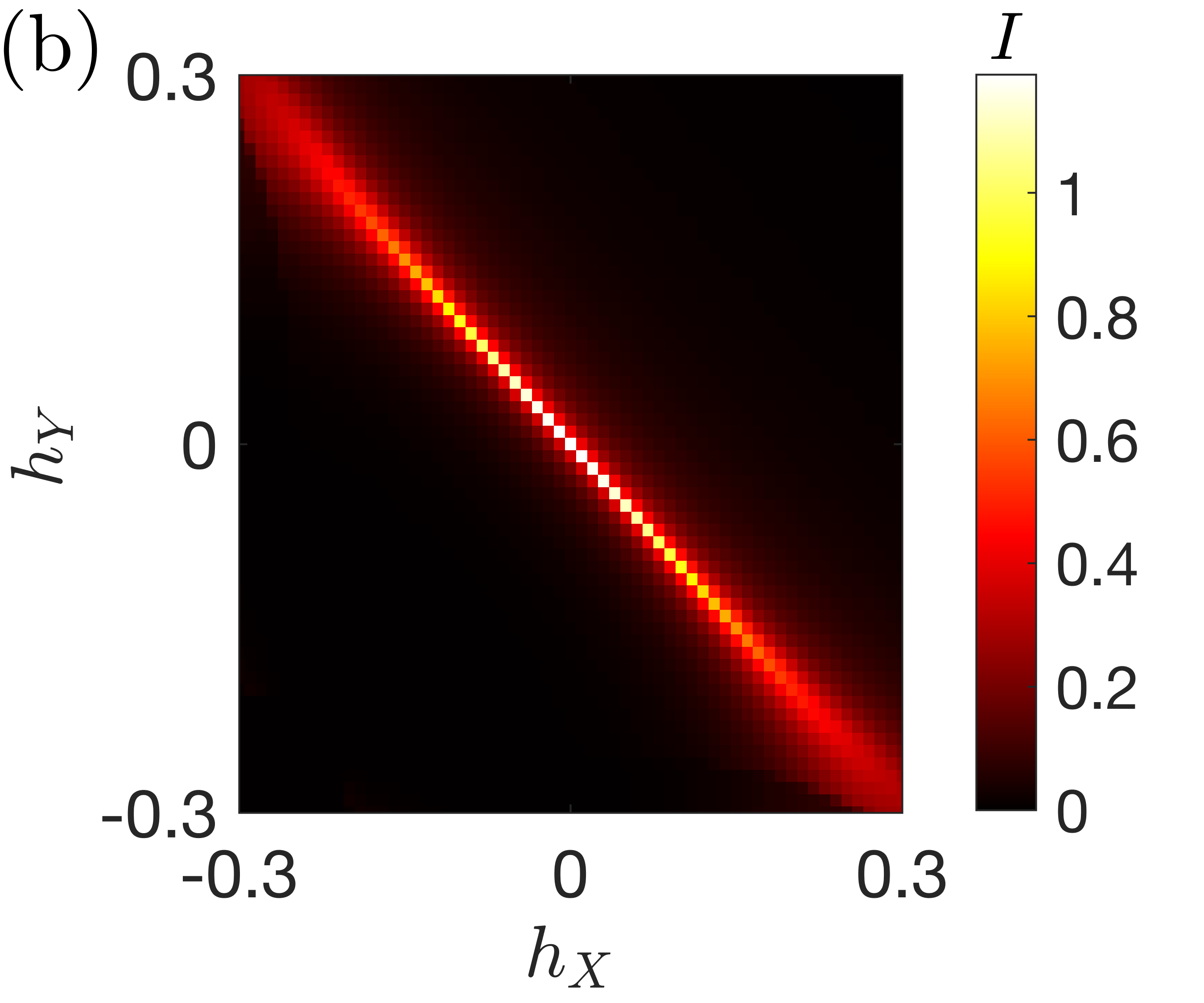} \\
    \includegraphics[width=0.25\columnwidth]{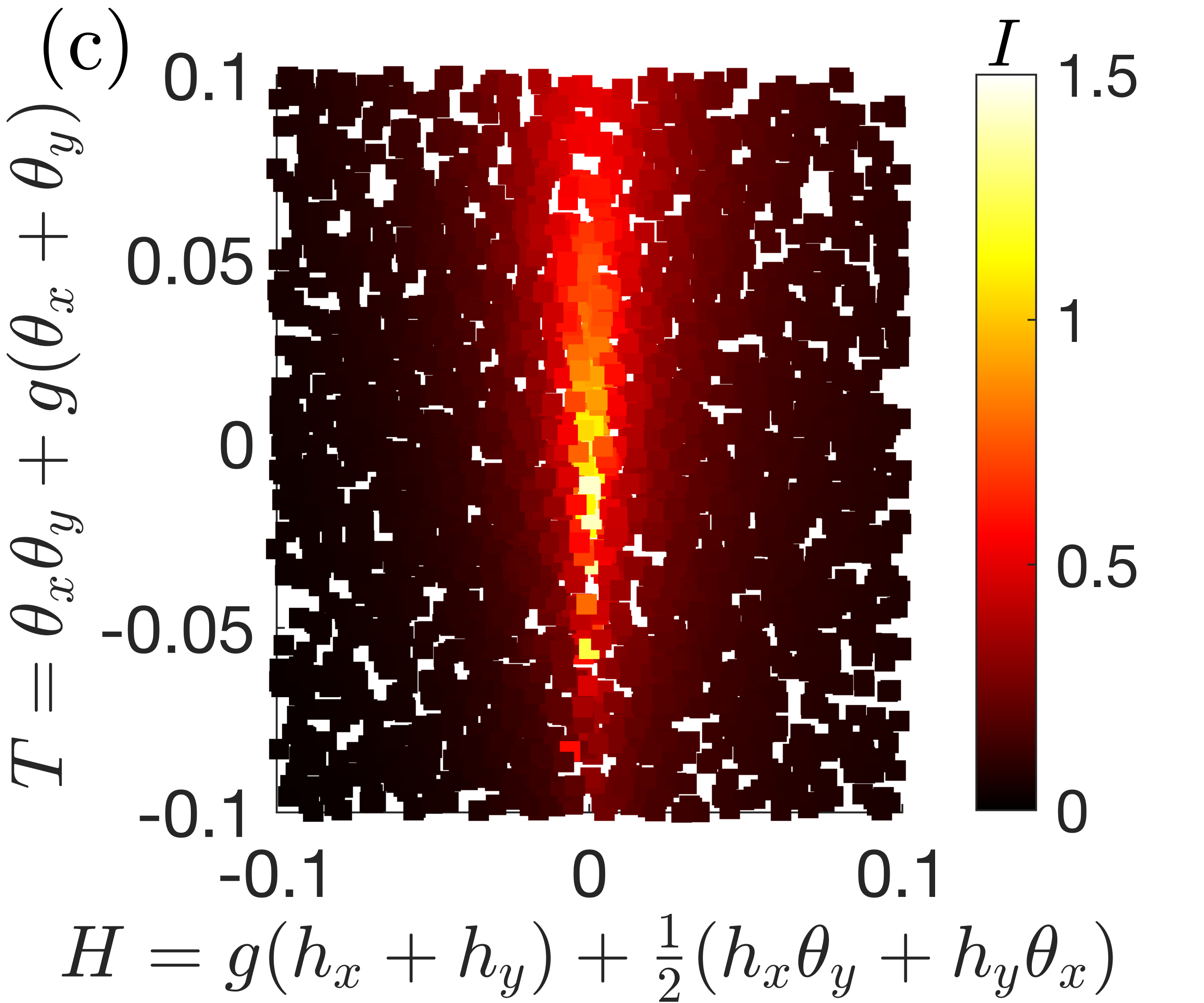}
    \includegraphics[width=0.25\columnwidth]{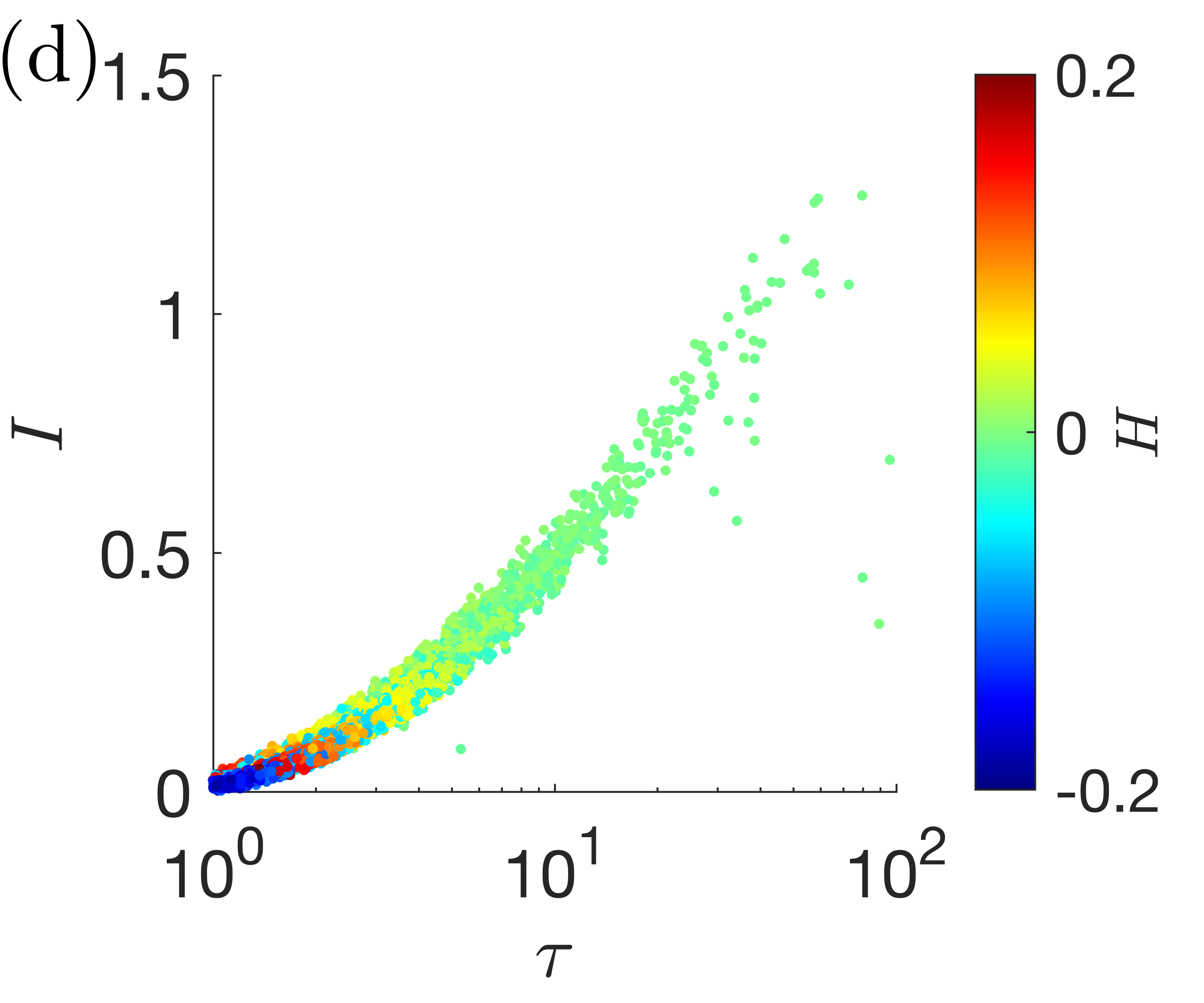}
	\caption{\r{\textbf{Shannon mutual information for dissimilar cells, using Hill feedback as in Eq.~\ref{eq:HillFeedback}}. \textbf{(a)} Dissimilar $\theta$ values with $h_X=h_Y=0$. The $T=0$ line is shown in dashed blue. \textbf{(b)} Dissimilar $h$ values with $\theta_X=\theta_Y=0$. \textbf{(c)} The mutual information $I$ vs.\ $H,T$ (Eq.~\ref{eq:HT}), generated from randomly uniformly drawn $h_x,h_y, \theta_x, \theta_y \in [-0.1, 0.1]$. \textbf{(d)} Mutual information as a function of the correlation time $\tau$ for the data shown in c. In all plots, $g=1$ and $n_c=3000$.}}
	\label{fig:Hill}
\end{figure}

\r{\section{Hill-function feedback}
This manuscript focuses on the steady-states of an extension of Schl\"ogl's second model \cite{schlogl1972chemical}. However, near the birfucation point a range of models can result in similar behavior \cite{Erez2019}. To verify this, we simulated a different realization of the birth/death dynamics, with feedback as a Hill function $f_j$: the birth rate $b_j=f_j$ and the death rate $d_j=j$. The Hill function has four parameters, $\{a, s, H, K\}$,
\be
	f_j = a + s\frac{j^H}{j^H+K^H}
	\label{eq:HillFeedback}
\ee
The mapping between the Hill function parameters and the Ising parameters for a single cell can be found in a previous manuscript \cite{Erez2019}. We detail the mapping for the two-cell Hill dynamics below. We simulated the Hill dynamics for the two cells similarly to Fig.~\ref{fig4}(a-b), shown in Fig.~\ref{fig:Hill}. Indeed, the same behavior emerges, with a peak of mutual information at the critical point and a high-information manifold at $T=0$ (dashed blue Fig.~\ref{fig:Hill}a),  and $h_X+h_Y=0$ (Fig.~\ref{fig:Hill}b).}

\r{\subsection{Derivation of the mapping between the Hill and Ising dynamics}
\noindent We consider the Hill feedback model with birth and death propensities,
\begin{gather}
		b_n = a_X + s_X \frac{n^H}{K + n^H}, \qquad d_n = k^- n,
\end{gather}
where $H>1$. For two cells, we have, 
\begin{align}
		\frac{dn_X}{dt} &= a_X + s_X \frac{n_X^{H_X}}{K_X^{H_X} +n_X^{H_X}} -k^- n_X -\gamma_{YX} n_X+\gamma_{XY} n_Y\,, \nn
		\frac{dn_Y}{dt} &= a_Y + s_Y \frac{n_Y^{H_Y}}{K_Y^{H_Y} +n_Y^{H_Y}} -q^- n_Y +\gamma_{YX} n_X-\gamma_{XY} n_Y\,. \nonumber
\end{align}
The second derivatives of the two equations vanish at
\begin{equation*}
	n_{cX} = K_X \left(\frac{H_X-1}{H_X+1}\right)^{1/H_X}\,, \mbox{ and similarly for $n_{cY}$}\,.
\end{equation*}
We specialize to the case where the Hill coefficients ($H_X, H_Y$) and half maximal values ($K_X,K_Y$) are the same, meaning that $H_X=H_Y=H$ and $K_X=K_Y=K$, which implies that $n_{cX}=n_{cY}$. Substituting $n_X = n_{cX}(m_X+1)$ and keeping terms to third order gives, 
	\begin{equation}\label{eq:m1Unscaled}
		\begin{aligned}
		n_c\frac{dm_X}{dt} &= \left[ -\frac{s_X}{2H} +a_X +\frac{s_X}{2} -(k^- +\gamma_{YX} -\gamma_{XY})n_c \right] +\gamma_{XY} n_c m_Y \\
		&+ n_c \left[ -k^- -\gamma_{YX} + \frac{(H^2-1)s_X}{4H n_c} \right] m_X - \frac{(H^2-1)^2 s_X}{48 H} m_X^3.
		\end{aligned}
	\end{equation}
	The analogous equation for $m_Y$ is
	\begin{equation}\label{eq:m2Unscaled}
	\begin{aligned}
	n_c\frac{dm_Y}{dt} &= \left[ -\frac{s_Y}{2H} +a_Y +\frac{s_Y}{2} -(q^- +\gamma_2 -\gamma_{YX})n_c \right] +\gamma_{YX} n_c m_X \\
	&+ n_c \left[ -q^- -\gamma_{XY} + \frac{(H^2-1)s_Y}{4H n_c} \right] m_Y - \frac{(H^2-1)^2 s_Y}{48 H} m_Y^3.
	\end{aligned}
	\end{equation}
To reach the Landau form, we rescale time by the factor $f$,
	\begin{equation}
		f = \frac{(H^2-1)^2 s_X}{16 H} \Longrightarrow \tau = \frac{f t}{n_c}.
	\end{equation}
Dividing Eq. \ref{eq:m1Unscaled} by the factor $f$ gives
	\begin{equation}\label{eq:m1Scaled}
	\begin{split}
	\frac{dm_X}{d\tau} = -\left[ \frac{8[s_X-H(2a_X +s_X-2n_c (k^- +\gamma_{YX}-\gamma_{XY}))]}{(H^2-1)^2 s_X } \right]  \\
	 + \left[ \frac{4((H^2-1) s_X - 4Hk^- n_c)}{(H^2-1)^2 s_X }\right] m_X - \frac{1}{3} m_X^3 \\
	+\frac{16H n_c}{(H^2-1)^2 s_X } (\gamma_{XY}m_Y -\gamma_{YX}m_X)\,.
	\end{split}
	\end{equation}
For the $m_Y$ equation, it will be convenient to introduce the ratio,
	\begin{equation}
		\tilde{\rho} = \frac{s_X}{s_Y}.
	\end{equation}
The term multiplying $m_Y^3$ in the equation for $dm_Y/dt$ is $-f/3\rho$. Multiplying Eq. \ref{eq:m2Unscaled} by $\rho/f$ gives
	\begin{equation}\label{eq:m2Scaled}
	\begin{split}
	\tilde{\rho}\frac{dm_Y}{d\tau} = -\left[ \frac{8[s_Y-H(2a_Y +s_Y-2n_c (q^- +\gamma_{XY}-\gamma_{YX}))]}{(H^2-1)^2 s_Y } \right]  \\
	+ \left[ \frac{4((H^2-1) s_Y - 4Hq^- n_c)}{(H^2-1)^2 s_Y }\right] m_Y - \frac{1}{3} m_Y^3 \\
    +\frac{16H n_c }{(H^2-1)^2 s_Y} (\gamma_{YX} m_X - \gamma_{XY} m_Y)\,.
	\end{split}
	\end{equation}	
Anticipating the Landau form, we define the mapping,
	\begin{equation}\label{eq:Hillparams}
		\begin{gathered}
		h_X = -\left[ \frac{8[s_X-H(2a_X +s_X-2n_c (k^- +\gamma_{YX}-\gamma_{XY}))]}{(H^2-1)^2 s_X } \right], \\
		h_Y =  -\left[ \frac{8[s_Y-H(2a_Y +s_Y-2n_c (q^- +\gamma_{XY}-\gamma_{YX}))]}{(H^2-1)^2 s_Y } \right], \\
		\theta_X = -\left[ \frac{4((H^2-1) s_X - 4Hk^- n_c)}{(H^2-1)^2 s_X }\right], \\
		 \theta_Y = -\left[ \frac{4((H^2-1) s_Y - 4Hq^- n_c)}{(H^2-1)^2 s_Y }\right], \\
		 g_{YX} =\frac{16Hn_c }{(H^2-1)^2 s_X}\gamma_{YX}, \qquad g_{XY} =\frac{16H n_c }{(H^2-1)^2 s_X }\gamma_{XY}\,.
		\end{gathered}
	\end{equation}	
We can now write the two-cell Hill dynamics in the Landau form,
	\begin{equation}
		\begin{gathered}
		\frac{dm_X}{d\tau} = h_X -\theta_X m_X -\frac{1}{3}m_X^3 -g_{YX}m_X + g_{XY}m_Y, \\
		\tilde{\rho}\frac{dm_Y}{d\tau} = h_Y -\theta_Y m_Y -\frac{1}{3}m_Y^3 +\tilde{\rho}g_{YX}m_X - \tilde{\rho}g_{XY}m_Y.
		\end{gathered}
\end{equation}
Now can now invert the expression in Eq. \ref{eq:Hillparams}. First, one specifies $n_c$ and $H$, this completely determines $K$. Further simplifying, $g_{YX}=g_{XY}=g$, we find,
\begin{equation}
	\begin{gathered}
	a_X = \frac{(H-1) \left((h_X+\theta_X)(H+1)^2+4\right)}{(H+1) \left(\theta_X \left(H^2-1\right)+4\right)}\,k^-n_c, \\
	a_Y = \frac{(H-1)  \left((h_Y+\theta_Y)(H+1)^2+4\right)}{(H+1) \left(\theta_Y\left(H^2-1\right)+4\right)}\,q^- n_c, \\
	s_X = \frac{16 H}{\left(H^2-1\right) \left(\theta_X \left(H^2-1\right)+4\right)}\,k^- n_c, \\
	s_Y= \frac{16 H}{\left(H^2-1\right) \left(\theta_Y \left(H^2-1\right)+4\right)}\, q^- n_c, \\	
	\gamma_{YX}= \frac{\left(H^2-1\right)}{\theta_X \left(H^2-1\right)+4}\,k^- g  =\gamma_{XY}\,.		
	\end{gathered}
\end{equation}
Importantly, the canonical Landau form (Eq.~\ref{eq:LandauForm}) requires that $\tilde{\rho}=1$. This dictates a relation between the degradation timescales $k^-$ and $q^-$ such that,
\be
\frac{q^-}{k^-} = \frac{\theta_Y(H^2-1)+4}{\theta_X(H^2-1)+4}
\ee
}

\r{\section{Dissimilar cells}
For dissimilar cells, we consider the mean molecule count as a function of the collective coordinates $H,T$,
\begin{figure}[!h]
	\includegraphics[width=0.25\columnwidth]{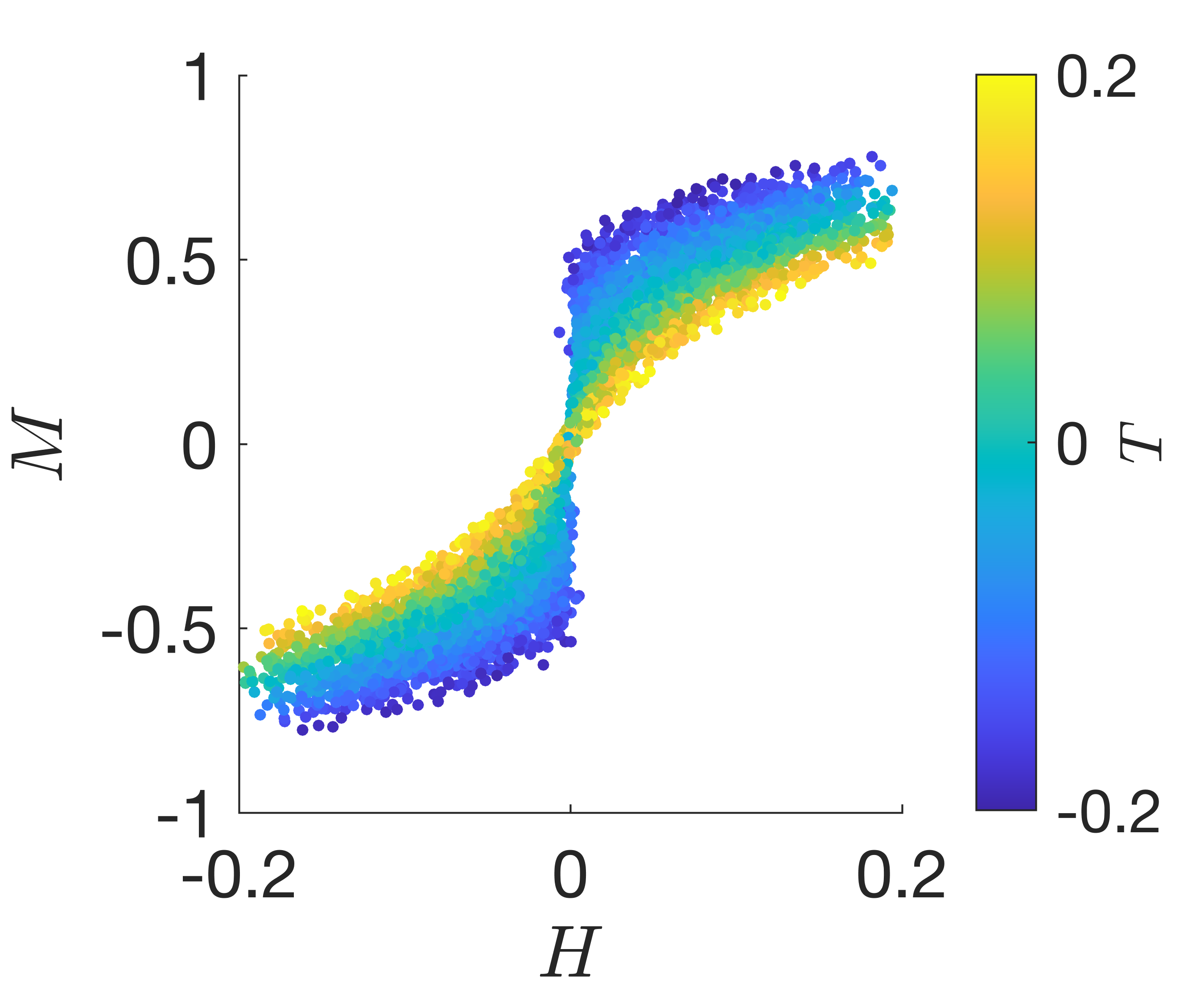}
	\includegraphics[width=0.25\columnwidth]{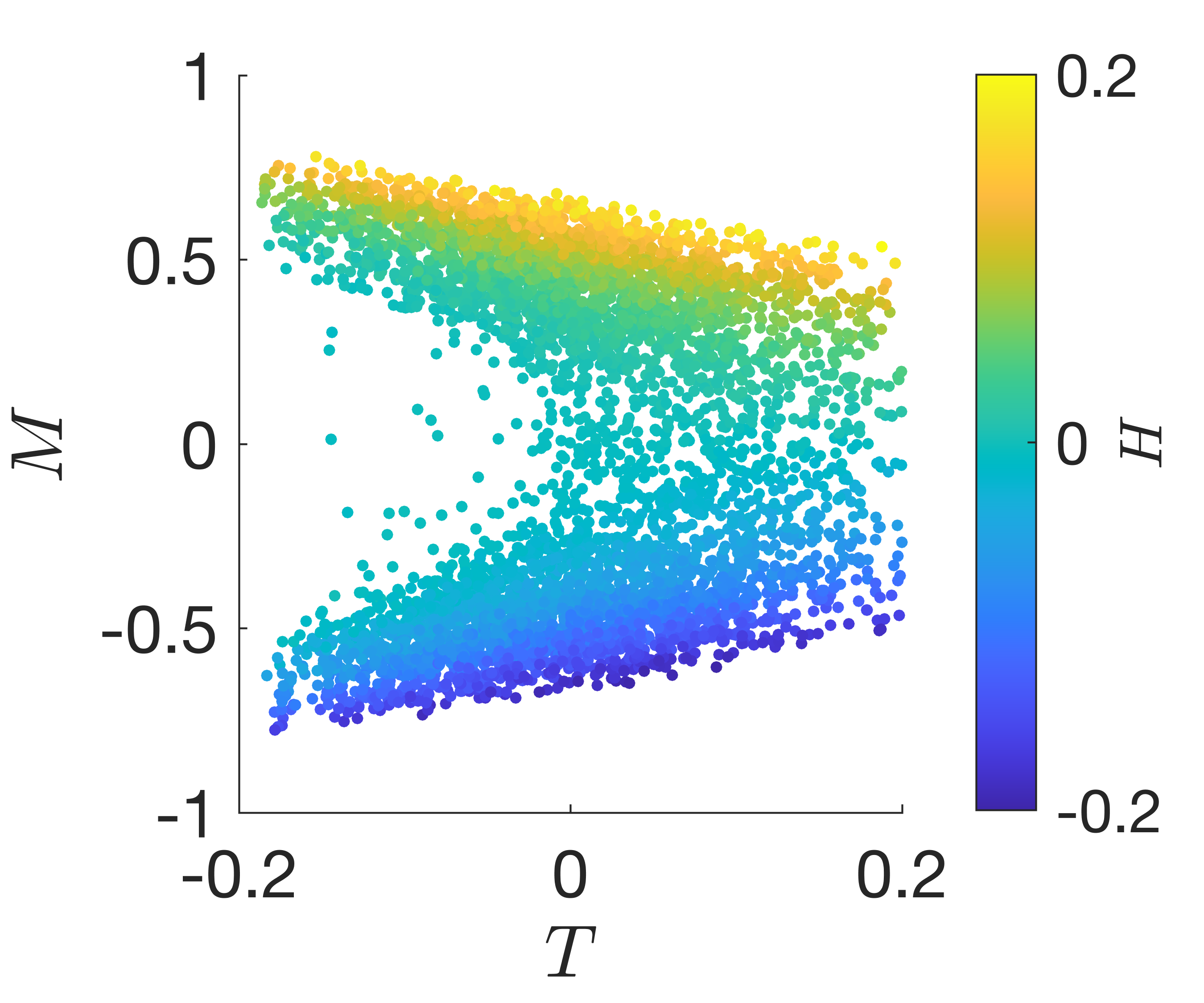}
	\caption{\r{\textbf{Mean molecule count $M$ as a function of $H$ and $T$}. We define a proxy for the joint magnetization, $M = \left(\frac{1}{2}(\bar{X}+\bar{Y})- n_c\right)/n_c$, with $\bar{X}$ the mean number of $X$ molecules. This allows us to consider the joint magnetization as a function of the collective coordinates $T, H$. \textbf{(a)} colored by $T$, \textbf{(b)} colored by $H$.}}
	\label{fig:MHT}
\end{figure}}
 
 \newpage
 
\section{Gaussian approximation for $h_X \ne h_Y$ case}
Fig.\ 4(b) of the main text shows a ridge in the mutual information when $h_X+h_Y=0$. Here, we approximate the joint molecule number distribution as Gaussian to understand the appearance of this ridge. For a pair of Gaussian random variables with covariance matrix $\mathcal{C}$, the mutual information is
	\begin{equation}\label{eq:gaussMI}
		I = \frac{1}{2} \log\left( \frac{\mathcal{C}_{xx} \mathcal{C}_{yy}}{\det \mathcal{C}} \right),
	\end{equation}
where $\det \mathcal{C} = \mathcal{C}_{xx}\mathcal{C}_{yy}-\mathcal{C}_{xy}^2$. We obtain the covariance matrix by writing down the Langevin equations corresponding to the reactions in Fig.\ 1(a). Specifically, we linearize the Langevin equations, which yields an Ornstein-Uhlenbeck process, whose steady state covariance matrix is known from It{\^o} calculus to take the form \cite{klebaner2012introduction}
	\begin{equation}\label{eq:covarIntegral}
		\mathcal{C} = \int_{0}^{\infty} e^{\mathbb{A}t} \mathbb{B} \mathbb{B}^T e^{\mathbb{A}^T t} dt,
	\end{equation}
where
\begin{equation}
\label{eq:mats}
	\mathbb{A} = \begin{bmatrix}
	-(c_X+\gamma) & \gamma \\
	\gamma & -(c_Y +\gamma)
	\end{bmatrix}, \qquad
	\mathbb{B} = \begin{bmatrix}
	\sqrt{\overline{b}_X} & -\sqrt{\overline{d}_X} & -\sqrt{\gamma \overline{x}} & 0 & 0 & \sqrt{\gamma \overline{y}} \\
	0 & 0 & -\sqrt{\gamma \overline{x}} & \sqrt{\overline{b}_Y} & -\sqrt{\overline{d}_Y} & -\sqrt{\gamma \overline{y}}
	\end{bmatrix}
\end{equation}
are the linearized Jacobian and Langevin noise matrices at the mean molecule numbers $\bar{x}$ and $\bar{y}$, written for simplicity in terms of the total birth and death rates and their derivative,
	\begin{equation}
	\label{eq:totProp}
		b_X(x) = k_1^+ + k_2^+ x^2, \qquad d_X(x) = k_1^- x + k_2^- x^3, \qquad c_X = \partial_x[d_X(x)-b_X(x)],
	\end{equation}
all evaluated at $\bar{x}$, and similarly for $Y$ (with $k\to q$).

We express the rates in terms of the Ising parameters using the mapping below Eq.\ \ref{eq:landau},
which in the limits of Fig.\ 4(d) ($n_{cX} = n_{cY} =n_c$, $\theta_X=\theta_Y=0$, $k_1^-=q_1^-$, $g_{XY} = g_{YX}=g$) simplify to
	\begin{equation}
	\label{eq:ratessimp}
		k_1^+ = k_1^- n_c (h_X+1/3), \qquad
		 k_2^+ =  \frac{k_1^-}{n_c}, \qquad
		k_2^- =  \frac{k_1^-}{3n_{c}^2}, \qquad
		\gamma =  k_1^- g.
	\end{equation}
We express the mean molecule numbers $\bar{x} = n_c(1+m_X)$ and $\bar{y} = n_c(1+m_Y)$ in terms of the Ising order parameters, which at steady state satisfy
	\begin{equation}
		\begin{gathered}
		0 = -\frac{1}{3} m_X^3 + h_X + g(m_Y - m_X), \\
		0 = -\frac{1}{3} m_Y^3 + h_Y + g(m_X - m_Y).
		\end{gathered}
	\end{equation}
These equations are solved by
	 	\begin{equation}
	\frac{1}{g^3}\left(\frac{m_X^9}{81}-\frac{h_X m_X^6}{9}+\frac{h_X^2 m_X^3}{3}-\frac{h_X^3}{3}\right)
	+ \frac{1}{g^2}\left(\frac{m_X^7}{9}-\frac{2 h_X m_X^4}{3}+h_X^2 m_X\right)
	+ \frac{1}{g}\left(\frac{m_X^5}{3}-h_Xm_X^2\right)
	+\frac{2 m_X^3}{3}-h_X-h_Y = 0.
	\end{equation}
and similarly for $X\leftrightarrow Y$. In the limit of small $g$ the first term dominates, and we recover the single-cell expectation $m_X = (3h_X)^{1/3}$. Conversely, in the limit of small $m$ and $h$ but order-one $g$ [as in Fig.\ 4(d)] the last term dominates, and we obtain $m_x = m_y = [3(h_X+h_Y)/2]^{1/3}$. Therefore
	\begin{equation}
	\label{eq:xy}
	\overline{x} = \overline{y} = n_c\left[1+ \left( \frac{3(h_X+h_Y)}{2} \right)^{1/3}\right].
	\end{equation}
	
Inserting Eq.\ \ref{eq:ratessimp} and \ref{eq:xy} into Eq.\ \ref{eq:totProp}, Eq.\ \ref{eq:totProp} into Eq.\ \ref{eq:mats}, and Eq.\ \ref{eq:mats} into Eq.\ \ref{eq:covarIntegral} yields analytic expressions for the elements of the covariance matrix $\mathcal{C}$. For small $h_X+h_Y$, the leading-order behavior of these elements is identical,
	\begin{equation}
	\mathcal{C}_{xx},\,\mathcal{C}_{yy},\,\mathcal{C}_{xy} \sim \frac{2 (12)^{1/3} n_c}{9(h_X +h_Y)^{2/3}}. 
	\end{equation}
This means that the numerator of Eq.\ \ref{eq:gaussMI} goes like $(h_X +h_Y)^{-2/3}$, whereas in the denominator, the leading-order terms cancel. Consequently, the numerator diverges more quickly than the denominator as $h_X+h_Y \rightarrow 0$, and therefore the mutual information diverges along this line.

We do not expect the Gaussian approximation to hold precisely at the critical point $h_X=h_Y=0$. Indeed, the mutual information does not diverge, but rather has a finite maximum near this point, i.e.\ the ridge in Fig.\ 4(b). Nonetheless, the divergence that we derive here provides an intuitive explanation for the ridge.


\end{document}